\documentstyle[12pt]{article}
\begin{document}
\baselineskip=14pt

\begin{center}
{\Large {\bf Parity Violation and Neutrino Mass}}

\vskip 0.3cm
            Tsao Chang \\ 
  CSPAR, University of Alabama, Huntsville, AL 35899, USA\\
   Hoff Lu Research Institute,  Shanghai, 200041, China\\
      Email: tsaochang@yahoo.com\\
\end{center}
{\bf Abstract }
Besides the fact of parity violation in weak interactions, 
based on evidences from neutrino oscillation and tritium 
beta decay, a natural conjecture is that neutrinos may be 
spacelike particles with a tiny proper mass.   A Dirac-type 
equation for spacelike neutrinos is further investigated 
and its solutions are discussed.  This equation can be written 
in two spinor equations coupled together via nonzero proper mass 
while respecting maximum parity violation.
\vskip 0.5cm
\noindent
{\bf Introduction}

Parity violation is a specific feature of the weak interactions. 
It was first discovered by T.D. Lee and C.N. Yang in 1956 [1] 
and experimentally established by C.S. Wu in beta-transition 
of polarized Cobalt nuclei [2]. In the standard model, neutrinos 
are massless.  Three flavors of neutrinos are purely 
left-handed, but anti-neutrinos are right-handed. In recent 
years, many evidences for neutrino oscillation 
come from the solar and atmospheric neutrino data have shown that 
neutrinos have tiny mass (about 1 eV) or mass difference [3-5].
  
If neutrino has a tiny rest mass, it would move slower than
light.  When taking a Lorentz boost with a speed
faster than the neutrino, the helicity of the neutrino would
change its sign in the new reference frame.  In another word,
parity would not be violated in weak interactions.
In order to solve this dilemma, The hypothesis that neutrinos 
may be spacelike particles is further investigated in this 
paper.     
  
Besides neutrino oscillations, recent measurements in 
tritium beta decay experiments have presented a value of negative 
mass squared,  $m^2(\nu_e)\approx -2.5 \  eV^2$ [6-8].  Moreover, 
the muon neutrino also exhibits a negative mass squared [9]. 
These results suggest that neutrinos may be tachyons.  
The negative value of the neutrino mass-square simply means:
     $$  E^2/c^2 - p^2 = m_{\nu}^2 c^2 < 0	\eqno  (1)  $$    
The right-hand side in Eq.(1) can be rewritten as ($-m_s^2 c^2$), 
then $m_s$ has a positive value. $m_s$ is called the proper mass or 
"meta mass" since it is different from the rest mass. 
The subscript $s$ means spacelike particle, i.e. tachyon.  
The negative value on the righthand side of Eq.(1) means that 
$p^2$ is greater than $(E/c)^2$ [10-12].

A Dirac-type equation for tachyons was investigated by Chodos et 
al. [13]. A form of the lagrangian density for tachyonic neutrinos 
was proposed.  More theoretical work can be found in Ref.[14-16]. 
In a recent paper, H-B. Ai has given a unified consideration for
neutrino oscillation and negative mass-square of neutrinos, which
should be paid more attention [17].\\
\noindent
{\bf Theory}

To follow Dirac's approach [18], the Hamiltonian form of spacelike 
Dirac-type equation for neutrinos can be written in:
$$  \hat E \Psi = c({\vec \alpha} \cdot {\hat p})\Psi + 
           \beta_s m_s c^2 \Psi 	     \eqno (2)  $$
with  ($\hat E = i\hbar \partial /\partial t , {\hat p} = 
-i \hbar \nabla $).  ${\vec \alpha} = (\alpha_1, \alpha_2, 
\alpha_3$) and $\beta_s$ are 4$\times$ 4 matrix, which are defined as
  $$ {\alpha_i} = \left(\matrix{0&\sigma_i\cr
                         \sigma_i&0\cr}\right),  \quad
   \beta_s = \left(\matrix{0&I\cr
                         -I&0\cr}\right)  \eqno (3)   $$
where $\sigma_i$ is 2$\times$2 Pauli matrix, $I$ is 2$\times$2 unit 
matrix.  The commutation relations are as follows:
 $$ \alpha_i \alpha_j + \alpha_j \alpha_i = 2 \delta_{ij}    $$
      $$	\alpha_i \beta_s + \beta_s \alpha_i =  0  $$
      $$       \beta_s^2 = -1	\eqno    (4)		$$
where $\beta_s = \beta \gamma_5$. The spacelike 
Dirac-type equation (2) can be rewritten in 
covariant forms by multiplying matrics $\beta$ and $\gamma_5$.  
Denote the bispinor function  $\Psi $ by two spinor functions:
$\varphi$ and $\chi$, then Eq.(2) can be rewritten as a pair of 
two-component equations:
$$ i\hbar \frac{\partial \varphi}{\partial t} = -ic 
    \hbar {\vec \sigma} \cdot  \nabla \chi + m_s c^2 \chi   $$
	$$ i\hbar \frac{\partial \chi}{\partial t} = -ic \hbar 
\vec{\sigma} \cdot \nabla \varphi - m_s c^2 \varphi  
\eqno (5)   $$
From Eq.(5), the conserved current can be derived:
$$  \frac{\partial \rho}{\partial t} +
           \nabla \cdot {\vec j} = 0  \eqno (6)   $$
and we obtain
  $$ \rho = \Psi^{\dag}  \gamma_5 \Psi ,  \quad
  {\vec j} = c(\Psi^{\dag} \gamma_5 {\vec \alpha} \Psi) 
     \eqno (7)  $$
where $\rho$ and $\vec j$ are probability density and current; 
$\Psi^{\dag}$ is the Hermitian adjoint of $\Psi$ .

Considering a plane wave along the $z$ axis for a right-handed 
particle, $\bar \nu$, the helicity $H = ({\vec \sigma} \cdot 
{\vec p})/|{\vec p}| = 1 $, then Eq.(5) yields the following 
relation:
  $$	\chi = \frac{cp - m_s c^2}{E} \varphi  \eqno	(8) $$
When taking massless limitation, $m_s = 0$, we obtain $E = cp$ and   
$\chi = \varphi$. Then Eq.(5) should be decoupled and reduced
to two-component Weyl equation [15-16].

The plane wave can be represented by
 $ \Psi(z,t)=\psi_{\sigma}exp[\frac{i}{\hbar}(pz-Et)] $
where $\psi_{\sigma}$ is a four-component bispinor.  Substituting
this bispinor into the wave equation (2) or (5), the explicit form
of two bispinors with positive-energy states are listed as follows:
$$ \psi_1=\psi_{\uparrow (+)} = N \left(\matrix{1\cr
     0\cr A \cr 0 \cr}\right), \quad
    \psi_2= \psi_{\downarrow (+)}  = N \left(\matrix{0\cr
                       -A \cr 0 \cr 1 \cr}\right)    \eqno (9) $$
and other two bispinors with the negative-energy states are:
$$ \psi_3=\psi_{\uparrow (-)} = N \left(\matrix{1\cr
     0 \cr -A \cr 0 \cr}\right), \quad
    \psi_4= \psi_{\downarrow (-)}  = N \left(\matrix{0\cr
                       A \cr 0 \cr 1 \cr}\right)    \eqno (10) $$
where the component $A$ and the normalization factor $N$ 
are chosen as
$$  A=\frac{cp-m_s c^2}{|E|},  \quad
    N=\sqrt{\frac{p+m_s c}{2m_s c}}        \eqno (11)  $$

For $ \psi_1=\psi_{\uparrow (+)}$, the conserved current in Eq.(6)
becomes:
$$ \rho = \Psi_1{^\dag} \gamma_5 \Psi_1=\frac{|E|}{m_s c^2}, \quad
     j = \frac{p}{m_s}     \eqno (12)  $$
Clearly, the ratio $j/\rho$ represents the superluminal speed $u_s$.
For $ \psi_2=\psi_{\downarrow (+)}$, the density $\rho$ is negative
so that it should be discarded.  If we consider the negative states 
as mathematics solutions in the preferred frame, then $\psi_1 = 
\psi_{\uparrow (+)}$ is the only solution with physical identity.  
It gives a natural choice that antineutrino is right-handed only.  
If we identify the preferred frame with Cosmic Microwave Background
Radiation (CMBR), the earth frame can be considered 
as the preferred frame approximately ($v/c \approx 10^{-3}$). 
Further study on the negative states and negative $\rho$ may be 
associated with complicated mathematics under Generalized Galilean
Transformation (GGT) (see [14]), which will not be discussed 
here. In addition, the pseudo scalar for each spinor satisfies:
${\bar\Psi} \gamma_5 \Psi = {\Psi}^+ \gamma^o \gamma_5 \Psi = 0 $.\\
 
\noindent
{\bf Discussion} 

It has been shown that the spacelike Dirac-type equation (2) 
reduces to the two component Weyl equation in the massless limit
[15, 16].  Notice that Eq.(2) is valid for the right-handed 
antineutrinos. In order to describe the left-handed neutrino, 
we now take a minus sign for the momentum operator, then Eq.(2) 
becomes
$$  \hat E \Psi_\nu = -c ({\vec \alpha} \cdot {\hat p})\Psi_\nu + 
           \beta_s m_s c^2 \Psi_\nu 	     \eqno (13)  $$
Similar to the solutions for Eq.(2), Eq.(13) yields one physical 
solution for the neutrino: $\psi_\nu = \psi_{\downarrow (+)}$.
Therefore, only $\bar \nu_R$ and $\nu_L$ exist in nature.

As is shown in [12, 19], the energy of a tachyonic neutrino (or 
anti-neutrino), $E_\nu$, could be negative in non-preferred 
reference frames. With respect to CMBR we obtain $\Delta E_{\infty} 
\approx 10^{-3}\ eV$ in the earth frame since the electron neutrino 
mass is about $1\ eV$. $\Delta E_{\infty}$ is a undetectable effect 
at present time. 

 The electron neutrino and the muon neutrino may have slightly
different proper masses.  It provides a natural explanation why 
the numbers of e-lepton and $\mu$-lepton are conserved respectively
at least for low energy experiments.

 Comparing with the electron mass, the mass term of the e-neutrino
in Eq.(2) is approximately close to zero.  Moreover, from the result
${\bar\Psi} \gamma_5 \Psi = 0$ for Spacelike neutrinos, it means 
that the mass term in Eq.(2) may be negligible in most experiments.  
Therefore, spacelike neutrinos behave just like the massless neutrinos. 
This similarity may also play role at the level of SU(2) gauge 
theory.  Indeed, the hypothesis of spacelike neutrinos  is 
a development of the two-component model of the massless 
neutrino, in which the reversal of its helicity is completely 
impossible.  On the other hand, if the neutrino has a small rest 
mass, it would be against the fact of parity violation in weak 
interactions.

 According to special relativity [20], if there is a spacelike 
particle, it might travel backward in time.  Besides the 
re-interpretation rule introduced in the 1960's [10,11], another 
approach is to introduce a kinematic time under GGT, which 
always goes forward [21-23].  Therefore, special relativity can 
be extended to the spacelike region without causality violation.\\

\noindent
{\bf Acknowlegement}
 
The author is grateful to G-j. Ni for helpful discussions. Thanks 
also due to helpful correspondence with H-B. Ai and E. Recami. \\

\end{document}